\documentclass[showpacs,showkeywords,preprintnumbers,amsmath,amssymb,pre,12pt]{revtex4}
\usepackage{amsmath,amssymb}
\usepackage{epsfig}
\usepackage{citesort}
\usepackage{graphicx}

\begin{document}
\title{A recipe for an unpredictable random number generator}
\author{M\'onica A. Garc\'{\i}a-\~Nustes\footnote{E--mail: mogarcia@ivic.ve}, Leonardo Trujillo\footnote{E--mail: leo@ivic.ve}
and Jorge A. Gonz\'alez\footnote{E--mail: jorge@ivic.ve} }
\affiliation{Centro de F\'{\i}sica, Instituto Venezolano de
Investigaciones Cient\'{\i}ficas (IVIC), \\
A.P. 21827, Caracas 1020--A, Venezuela}
\date{\today}
%
\begin{abstract}
 In this work we present a model for computation of random processes in
digital computers which solves the problem of periodic sequences and
hidden errors produced by correlations. We show that systems with
non-invertible non-linearities can produce unpredictable sequences
of independent random numbers. We illustrate our result with some
numerical calculations related with random walks simulations.
\end{abstract}
\maketitle

Many challenging problems in computational physics are associated
with reliable realizations of randomness (e.g Monte Carlo
simulations). In a typical 32-bit format a maximum of $2^{32}$
floating points numbers  can be represented. Therefore, a recursive
function $X_{n+1}=f(X_{n},X_{n-1},\ldots, X_{n-r+1})$ acting on
these numbers generates a sequence $X_{0},X_{1},X_{2},...,X_{N-1}$
which must repeat itself. It is known that for any recursive
function, a digital computer can only generate periodic sequences of
numbers \cite{Fer,Grass,Vatt,Souza,Nogues,Ecu,Mertens1}. These
generators are not unpredictable.

Definition of truly unpredictable: The next values are not
determined by the previous values. A process $X_n=P(\theta T Z^n)$
is said to be unpredictable if for any string of values
$X_{0},X_{1},X_{2},...,X_{m}$ of length $m+1$, generated using some
$\theta = \theta_1$, there are other values of $\theta$ for which
function $X_n=P(\theta T Z^n)$ generates exactly the same string of
numbers $X_{0},X_{1},X_{2},...,X_{m}$, but the next value $X_{m+1}$
is different, where $m$ is any integer. Note that this kind of
process cannot be expressed as a map of type
$X_{n+1}=f(X_n,X_{n-1},\ldots,X_{n-r+1})$
\cite{GroupPhysLettA,GroupPhysicaD,GroupEuroPL}




All known generators (in some specific physical calculations) give
rise to incorrect results because they deviate from randomness
\cite{Grass,Souza,Nogues,Mertens1}. It is trivial that any periodic
process is not unpredictable. Suppose $m_T$ is the period of the
generated sequence. Given any string of $m_T$ values:
$X_{s},X_{s+1},...,X_{m_T-1}$; the next value $X_{m_T}$ is always
known, because the process is periodic. On the other hand, for any
generator of type $X_{n+1}=f(X_n,X_{n-1},\ldots,X_{n-r+1})$, given
any string of $r$ values: $X_{s},X_{s+1},...,X_{s+r-1}$; the next
value $X_{s+r}$ is always determined by the previous $r$ values.
Thus it is not unpredictable. So the subsequences must be
correlated.

An example of this can be found in Ref.\cite{Nogues}, where the
authors have shown that using common pseudo random number
generators, the produced random walks present symmetries, meaning
that the generated numbers are not independent. On the other hand,
the logarithmic plot of the mean distance $<d>$ versus the number of
steps $N$ is not a straight line (as expected theoretically,
$<d>\sim N^{1/2}$) after $N>10^{5}$ (in fact, it is a rapidly
decaying function). Here $d$ is defined as the end to end
mean-square distance from the origin of the random walk as a
function of the number of steps. Other papers on the influence of
the pseudorandom number generator on random walk simulations are the
following \cite{Grass93,Shchur97,Shchur99}


In the following, we will show that using non-invertible nonlinear
functions, we can create an unpredictable random number generator
which does not contain visible correlations while simulating a
random walk with the length $10^9$.

Let us investigate the following
function\cite{GroupPhysLettA,GroupPhysicaD,GroupEuroPL}:
\begin{equation}\label{Pfunction}
    X_{n}=P(\theta \; T Z^{n}),
\end{equation}
where $P(t)$ is a periodic function, $\theta$ is a real number, $T$
is the period of the function $P(t)$, and $Z$ is a noninteger real
number.

Let $Z$ be a rational number expressed as $Z=p/q$, where $p$ and $q$
are relative prime numbers. Now let us define the following family
of sequences
\begin{equation}
X_{n}^{(k,m,s)}=P \left[ T (\theta_{0} + q^{m}k)\left(
\frac{q}{p}\right)^{s}\left( \frac{p}{q}\right)^{n}\right],
\end{equation}
where $k$, $m$ and $s$ are non-negative integers. The parameter $k$
distinguishes the different sequences. For all sequences
parametrized by $k$, the strings of $m+1$ values
$X_{s},X_{s+1},X_{s+2},...,X_{s+m}$ are the same. This is so because
\begin{equation}
\nonumber X_{n}^{(k,m,s)}=P\left[T \theta_{0}
\left(\frac{q}{p}\right)^{s} \left(\frac{p}{q}\right)^{n} + T k
p^{n-s}q^{(m-n+s)} \right] =P\left[T \theta_{0}
\left(\frac{q}{p}\right)^{s} \left(\frac{p}{q}\right)^{n}\right]
\end{equation}
%
for all $s\leq n \leq m+s$. Note that the number
$k\;p^{(n-s)}q^{(m-n+s)}$ is an integer for $s\leq n\leq m+s$. So we
can have an infinite number of sequences that share the same string
of $m+1$ values.

Nevertheless, the next value
$X_{s+1}^{(k,m,s)}= P \left[ T \; \theta_{0}
\left(\frac{q}{p}\right)^{s} \left(\frac{p}{q}\right)^{(s+1)} +
\frac{T\; k p^{(m+1)}}{q} \right]$
is uncertain. In general $X_{s+1}^{(k,m,s)}$ can take $q$ different
values. In addition, the value $X_{s-1}^{(k,m,s)}$,
$ X_{s-1}^{(k,m,s)}= P \left[ T \; \theta_{0}
\left(\frac{q}{p}\right)^{s} \left(\frac{p}{q}\right)^{(s-1)} +
\frac{T\; k q^{(m+1)}}{p} \right]$,
is also undetermined from the values of the string
$X_{s},X_{s+1},X_{s+2},...,X_{s+m}$. There can be $p$ different
possible values.   In the case of a generic irrational $Z$, there
are infinite possibilities for the future and the past. From the
observation of the string $X_{s},X_{s+1},X_{s+2},...,X_{s+m}$, there
is no method for determining the next and the previous  values of
the sequence.

But this is not the only feature of these functions. It can be
demonstrated that there are no statistical correlations between
$X_{m}$ and $X_{n}$ if $m\neq n$, and that they are also independent
in the sense that their probability densities satisfy the
relationship $P(X_{n},X_{m})=P(X_{n})P(X_{m})$
\cite{GroupPolonica,GroupJap}.

Moreover, we will show that, given the function (\ref{Pfunction}),
any string of sequences $X_{s},X_{s+1},\ldots, X_{s+r}$ constitutes
a set of statistically independent random variables.

Without loss of generality, we assume that $P(t)$ has zero mean and
can be expressed using the following Fourier representation
%
$ P(t)=\sum_{k=-\infty}^{\infty}a_k e^{i\pi k t}. $
%
%

We can calculate the $r$--order correlation functions
\cite{GroupPolonica,GroupJap}:
\begin{eqnarray}
\label{EqA1}
E(X_{n_1}\cdots X_{n_r})&=&\int_X d\theta P(T\theta    Z^{n_1})\cdots P(T\theta Z^{n_r})\nonumber\\
&=&\sum_{k_1=-\infty}^{\infty}\cdots\sum_{k_r=-\infty}^{\infty}a_{k_1}\cdots a_{k_r}\int_{0}^{1}d\theta \exp\left\{ i\pi(k_1Z^{n_1}+\cdots +k_rZ^{n_r})T\theta\right\}\nonumber\\
&=&\sum_{k_1=-\infty}^{\infty}\cdots\sum_{k_r=-\infty}^{\infty}a_{k_1}\cdots
a_{k_r}\delta(k_1Z^{n_1}+\cdots +k_rZ^{n_r},0),
\end{eqnarray}
where the coefficients $k_{i}$ can be different integers, and
$\delta (n,m)=1$ if $n=m$ or $\delta(n,m)= 0$ if $n\neq m$.

When all $n_i$ are even, the following equation is satisfied
\begin{equation}
\label{EE}
E(X_{s}^{n_1}X_{s+1}^{n_2}\cdots
X_{s+r}^{n_r})=E(X_{s}^{n_1})E(X_{s+1}^{n_{2}})\cdots
E(X_{s+r}^{n_{r}}).
\end{equation}

The main problem in this equation occurs when one of the numbers
$n_i$ is odd. In this case, the correlations $E(X_{s}^{n_1}
X_{s+1}^{n_2}\cdots X_{s+r}^{n_r})$ must be zero. A nonzero
correlation in Eq.(\ref{EE}) exists only for the sets
$(n_1,n_2,\ldots,n_r)$ that satisfy the equation $k_1Z^{n_{1}} +
\cdots + k_{r}Z^{n_{r}}=0$. For a typical real number $Z$, this
equation is never satisfied.


If we use non-invertible nonlinear functions, type of
(\ref{Pfunction}), we can implement a Truly Random Number Generator
(TRNG). In this case, we propose the following function
\begin{equation}
\label{generatingfunction} X_{n}=[\theta_{s} Z^{n}] \quad mod \;1
\end{equation}

Function (\ref{generatingfunction}) is an example of the general
case $X_n=P[\theta T Z^n]$ studied in this paper. We have shown that
the subsequences $X_s,X_{s+1},\ldots , X_{s+r}$ constitutes a set of
statistically independent random variables. The particular case of
function (\ref{generatingfunction}) is well--known to produce
uniformly distributed numbers
\cite{GroupPhysLettA,GroupPhysicaD,GroupEuroPL}.

Now we will formulate a central limit theorem. Using theorems proved
in previous studies \cite{GroupPolonica,GroupJap,r16,r17,r18,r19}
and the results obtained from this paper, we obtain the following
formula: If $Z$ is a generic real number and $X_n=2(Y_n-1/2)$,
$Y_n=[\theta Z^n]mod 1$, then

\begin{equation}\label{Eq30}
    \lim_{r\rightarrow\infty}P\left\{ \alpha<\frac{X_1+X_2+\cdots X_r}{\sqrt{r}}<\beta
    \right\}=\frac{1}{\sqrt{\pi}}\int_{\alpha}^{\beta}\mathrm{e}^{-\xi^2}\mathrm{d}\xi.
\end{equation}

The Gaussian distribution of the sums is correct even for other
functions $X_n=P[\theta Z^n]$, where $P(t)$ is periodic. This has
been shown in numerical simulations \cite{GroupPolonica}.

The numbers $X_n=[\theta Z^n]\mod 1$ are uniformly
distributed\cite{GroupPhysLettA,GroupPhysicaD,GroupEuroPL}. We can
simulate different stochastic processes (with different
distributions) using different functions $X_n=P[\theta T Z^n]$. As
$\rho (X_n)=1$, $\rho(X_{n+1})=1$, $\rho(X_{n},X_{n+1})=1$, it is
trivial that they are independent.

It is interesting to check the theoretical predictions using
numerical simulations of the behavior of different stochastic
processes.

For instance, let us study the function
\begin{equation}\label{EqU}
    U_n = \cos[2\pi\theta Z^n].
\end{equation}

All the moments and higher--order correlations can be calculated
exactly \cite{GroupPolonica,GroupJap}:

For odd m:
\begin{equation}\label{Eq15}
    E(U_{n}^{m})=0.
\end{equation}

If any $n_i$ is odd, then
\begin{equation}\label{Eq16}
    E(U_{s}^{n_0}U_{s+1}^{n_1}\cdots U_{s+r}^{n_r})=0
\end{equation}

Suppose now that all $n_i$ are even:
\begin{eqnarray}\label{Eq18}
 E(U_s^{n_0}U_{s+1}^{n_1}\cdots U_{s+r}^{n_r})= 2^{-(n_0+n_1+\cdots +n_r)} \left(
\begin{array}{c}
  n_0 \\
  \frac{n_0}{2} \\
\end{array}
\right) \left(
\begin{array}{c}
  n_1 \\
  \frac{n_1}{2} \\
\end{array}
\right)
\cdots
\left(
\begin{array}{c}
  n_r \\
  \frac{n_r}{2} \\
\end{array}
\right),
\end{eqnarray}
\begin{eqnarray}\label{Eq19}
 E(U_s^{n_0})= 2^{-n_0} \left(
\begin{array}{c}
  n_0 \\
  \frac{n_0}{2} \\
\end{array}
\right),
\end{eqnarray}
\begin{eqnarray}\label{Eq20}
 E(U_{s+1}^{n_1})= 2^{-n_1} \left(
\begin{array}{c}
  n_1 \\
  \frac{n_1}{2} \\
\end{array}
\right),\ldots,
\end{eqnarray}
\begin{eqnarray}\label{Eq21}
 E(U_{s+r}^{n_r})= 2^{-n_r} \left(
\begin{array}{c}
  n_r \\
  \frac{n_r}{2} \\
\end{array}
\right).
\end{eqnarray}

Note that the condition for independence is satisfied
\begin{equation}\label{Eq22}
    E(U_{s}^{n_0}U_{s+1}^{n_1}\cdots
    U_{s+r}^{n_r})=E(U_{s}^{n_0})E(U_{s+1}^{n_1})\cdots
    E(U_{s+r}^{n_r}),
\end{equation}
for all integer $n_{0},n_{1},\ldots n_{r}$.

We have performed extensive numerical simulations that confirm the
values of these moments and the independent conditions.

An additional checking is the following.

The probability density of $U_n$ is
$\rho(U)=\frac{1}{\pi\sqrt{1-U^2}}$. Define $V_n=U_{n+1}$. The
probability density of $V_n$ is $\rho(V)=\frac{1}{\pi\sqrt{1-V^2}}$.
We have checked both theoretically and numerically that
$\rho(U,V)=\frac{1}{\pi^2\sqrt{(1-U^2)(1-V^2)}}$, that is
$\rho(U,V)=\rho(U)\rho(V)$. This can be observed in Fig. 1 and Fig.
2.
\begin{figure}[ht]
\centerline{\includegraphics[height=6.0cm]{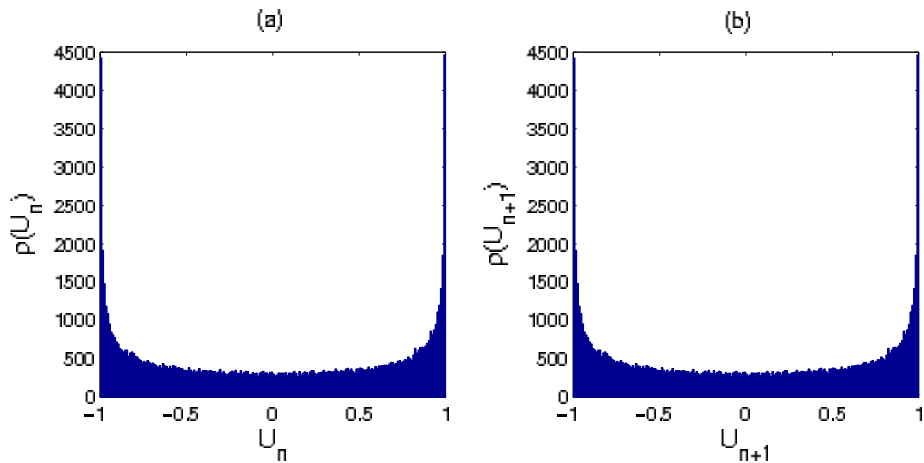}\hspace{0.5cm}}
\caption{Probability densities for random variables
$U_{n}=\cos[2\pi\theta Z^{n}]$ and $V_{n}=U_{n+1}$.
a)$\rho(U)=\frac{1}{\pi \sqrt{1-U^{2}}}$; b)$\rho(V)=\frac{1}{\pi
\sqrt{1-V^{2}}}$}
\end{figure}
\begin{figure}[ht]
\centerline{\includegraphics[height=6.0cm]{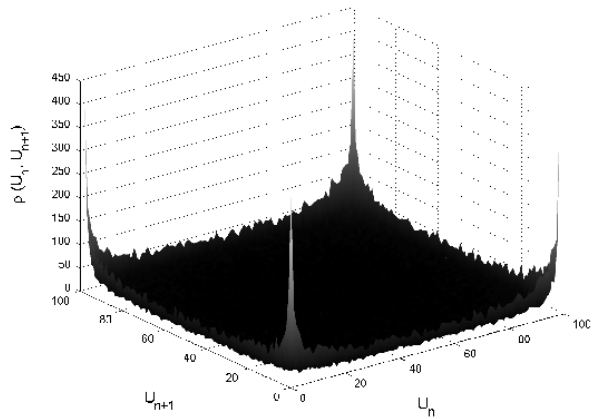}\hspace{0.5cm}}
\caption{Probability density $\rho(U_{n},U_{n+1})$, when
$U_{n}=\cos[2\pi\theta Z^{n}]$. Here $\rho(U,V)=\frac{1}{\pi^{2}
\sqrt{(1-U^{2})(1-V^{2})}}$}
\end{figure}
In order to avoid computation problems, we have used the following
procedure.
%
We change parameter $\theta$ after each set of $M$ values of
$X_{n}$, where $M$ is the maximum number for which there are not
overflow problems, in such a way that the next value of $X_{n+1}$ is
obtained with the new $\theta$. Let us define
\begin{equation}
\theta_{s}=A (C_{s}+X_{s}) + 0.1
\end{equation}
where $C_{s}$ is a  sequence obtained using the digits of the
Champernowne's number \cite{Champ} (i.e., $0.1234567891011...$):
$C_{0}=0.123456$ , $C_{1}=0.234567$, $C_{2}=0.345678$,
$C_{3}=0.456789$, $C_{4}=0.567891$, and so on. This sequence is
nonperiodic. Index $s$ is the order number of $\theta$ in such a way
that $s=0$ corresponds to the $\theta$ used for the first set of $M$
sequence values $X_{1}, X_{2},...,X_{M}$; $s=1$ for the second set
$X_{M+1}, X_{M+2},...,X_{2M}$, and so on. $X_{0}$ represents the
TRNG's seed.

Using this method we have generated a very long sequence of random
numbers without computational problems.


To test function (\ref{generatingfunction}) as a truly random number
generator, we have implemented a random walk simulation program in
C++. We have made a sampling test of a random walk with $N=10^{9}$
steps with $100$ realizations with different initial seeds. The mean
distance $\left<d\right>$ was calculated every $1000$ steps of the
random walk.

The Champernowne sequence of numbers used in the generator  was
produced previously by a short C++ program, who created a sequence
of a maximum of 40000 Champernowne's numbers. If a larger amount of
values to $C_{s}$ is necessary, it can be obtained using a segment
code that uses the 40 thousand values already stored in $C_{s}$ and
mixing them, e.g the algorithm takes the first value of the series
$C_{1}$,  the third $C_{3}$ and so on, and adds them at the end of
the series, obtaining that $C_{s+1}=C_{1}$, $C_{s+2}=C_{3}$,...; if
more values are necessary, this procedure or cycle is repeated but
now skipping two values $C_{1}$, $C_{3}$, $C_{5}$,... three values
$C_{1}$, $C_{4}$, $C_{7}$ and so on. In this way, we can make the
$C_{s}$ sequence as large as we wish.


We present  a logarithmic plot of the mean distance $<d>$ versus the
number of steps $N$ with $N=10^{9}$ steps with $A=6.9109366$ and
$Z=\pi/2$ (See Fig. 3). We can verify that there is no deviation
from the theoretical straight line, even for $N>> 10^{5}$ steps,
which is a very good test of the reliability of the Random Number
Generator used in the random walk simulations.

\begin{figure}[ht]
\begin{center}
\includegraphics[height=5.5cm]{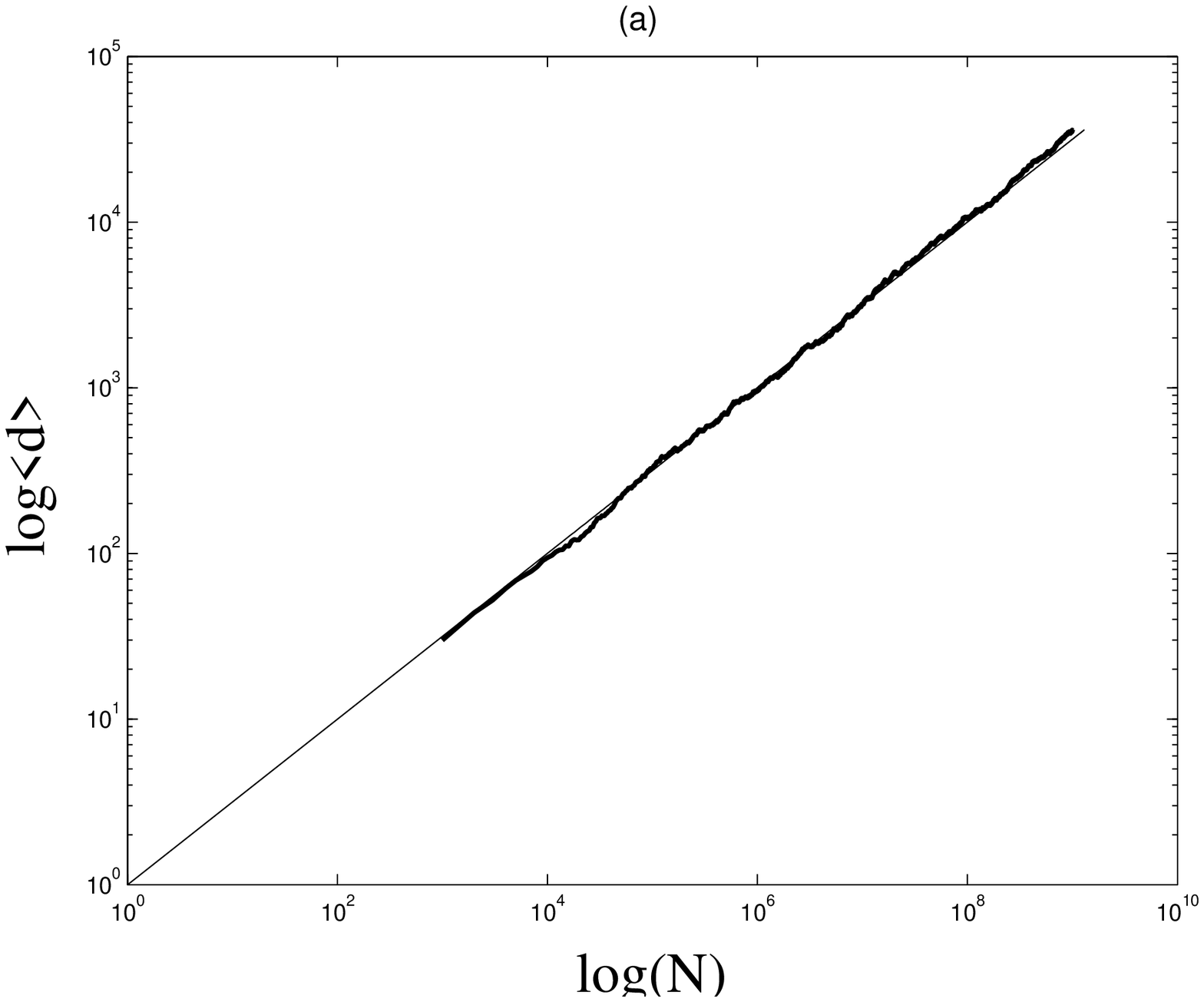}\hspace{0.5cm}
\includegraphics[height=5.5cm]{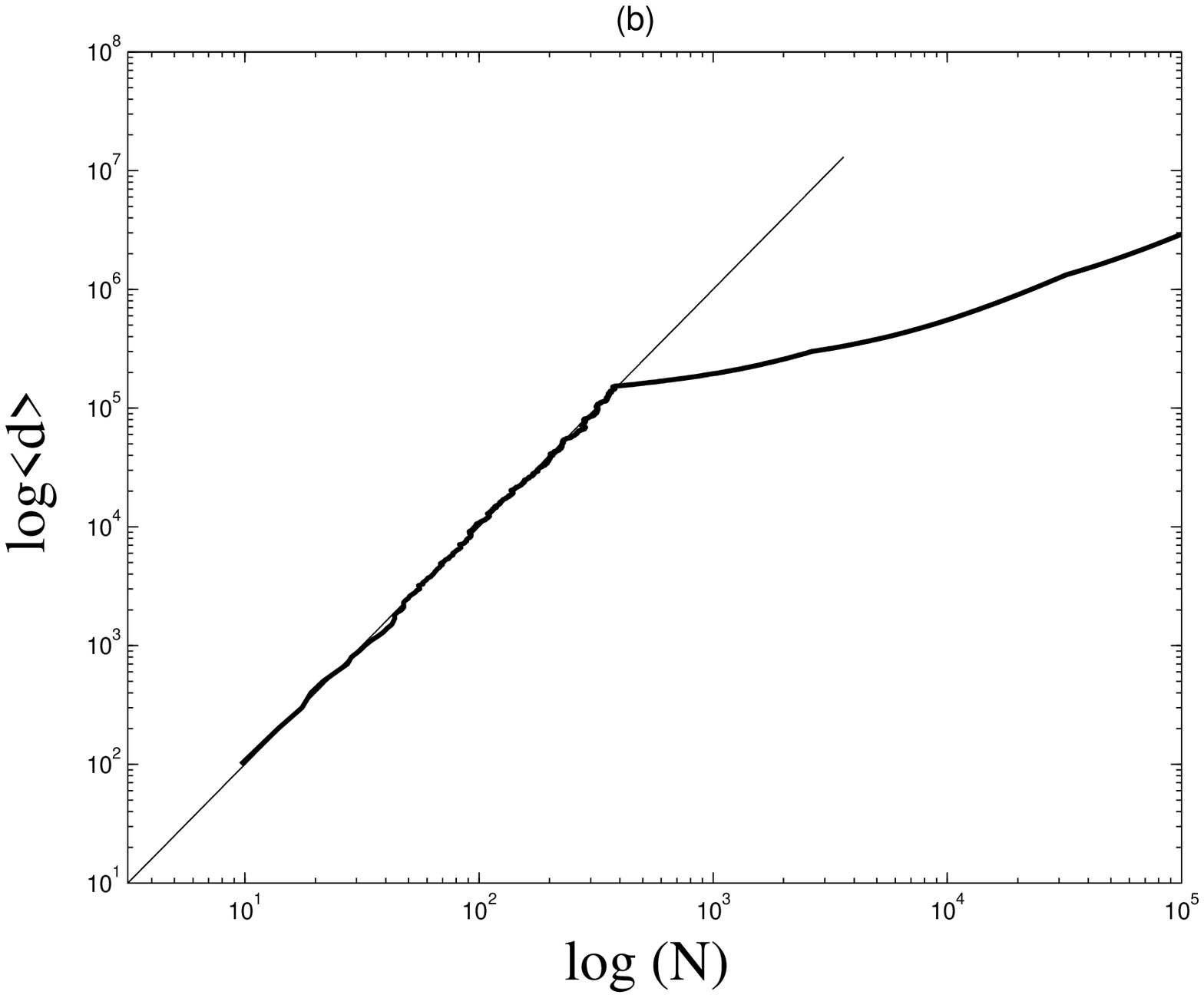}\hspace{0.5cm}
\end{center}
\caption{Logarithmic plot of the mean distance $<d>$ versus the
number of steps $N=10^{9}$ steps. a) for generator
(\ref{generatingfunction}), b) the same simulation for a generator
of type $X_{n+1}=aX_n\mod T$.}
\end{figure}


We have presented a random number generator based on the properties
of non--invertible transformations of truncated exponential
functions. The obtained random process is unpredictable in the sense
that the next values are not determined by the previous values. We
have applied this generator to the numerical simulation of
statistically independent random variables. In the simulation of a
random walk with the length $10^9$, the random process does not
contain visible correlations.


\begin{thebibliography}{2}
    \bibitem{Fer}A.M. Ferrenberg, D. P. Landau, Y.J. Wong, Phys. Rev.
    Lett. 69 (1992) 3382
    \bibitem{Grass}
    P. Grassberger, Phys. Lett. A 181 (1993) 43
    \bibitem{Vatt}I. Vattulainen, T. Ala--Nissila, K. Kankaala,
    Phys. Rev. Lett. 73 (1994) 2513
    \bibitem{Souza}R.M. D'Souza, Y.Bar-Yam, M. Kardar, Phys. Rev.E
    57 (1998) 5044
    \bibitem{Nogues}J.Nogu\'es, J.L. Costa-Kr\"{a}mer, K.V. Rao,
    Physica A 250 (1998) 327
    \bibitem{Ecu} P. L'Ecuyer, Oper. Res. 47 (1999) 159
    \bibitem{Mertens1}H. Bauke, S. Mertens, J. Stat. Phys. 114
    (2004) 1149
    055702
    \bibitem{GroupPhysLettA}
    J.A. Gonz\'alez, L.I. Reyes, J.J. Su\'arez, L.E. Guerrero, G.
    Guti\'errez, Phys. Lett. A 295 (2002) 25
    \bibitem{GroupPhysicaD}
    J.A. Gonz\'alez, L.I. Reyes, J.J. Su\'arez, L.E. Guerrero, G.
    Guti\'errez, Physica D 178 (2003) 26
    \bibitem{GroupEuroPL}
    L. Trujillo, J.J. Su\'arez, J.A. Gonz\'alez, Europhys. Lett. 66
    (2004) 638
    \bibitem{Grass93}
    P. Grassberger, J. Phys. A: Math. Gen. 26 (1993) 2769
    \bibitem{Shchur97}
    L.N. Shchur, J.R. Heringa and H.W.J. Bl\"{o}te, Physica A 241 (1997)
    579
    \bibitem{Shchur99}
    L.N. Shchur, Comput. Phys. Comm. 121 (1999) 83
    \bibitem{GroupPolonica}J.A. Gonz\'alez, L. Trujillo, Acta Physica Pol.
    B 37 (2005) 2394
    \bibitem{GroupJap}J.A. Gonz\'alez, L. Trujillo, J. Phys. Soc. Japan
    75 (2006) 023002
    \bibitem{r16}
    M. Kac,
    %
    Stud. Mathematica  6 (1936) 46
    \bibitem{r17}
    M. Kac and H. Steinhaus,
    %
    Stud. Mathematica 6 (1936) 59
    \bibitem{r18}
    M. Kac and H. Steinhaus,
    %
    Stud. Mathematica 6 (1936) 89
    \bibitem{r19}
    M. Kac and H. Steinhaus,
    %
    Stud. Mathematica 7 (1937) 1
    \bibitem{Champ}D.G Champernowne, J. London Math. Soc. 8 (1933)
    254


\end{thebibliography}
\end{document}